# Elite Proxies, Algorithmic Bottlenecks, and Multi-Platform Information Cascades in the 2026 Iran War

**Hakan Mehmetcik**
Visiting Researcher, Kellogg Institute for International Studies, University of Notre Dame, IN, USA; Prof of International Relations, at the Faculty of Political Science, Marmara University, Istanbul, Türkiye
hakan.mehmetcik@marmara.edu.tr
ORCID: 0000-0002-1882-4003

**Abstract**

As high-stakes executive crisis discourse shifts to fragmented digital ecosystems, unmediated narratives increasingly originate in closed-broadcast enclaves before spreading across structurally distinct open-networks. Yet the mechanics of cross-platform information diffusion remain under-theorized, necessitating rigorous computational investigation. This study traces the mutation of President Trump's statements during the 2026 Iran War as they migrate from Truth Social to X-Twitter and Bluesky. Utilizing a novel Metric-to-Semantic-Linkage framework and a high-throughput dataset of 9,891 temporally synchronized public response records, we examine how divergent platform architectures condition the downstream vernacular of geopolitical conflict. Our findings reveal a profound behavioral divergence driven by sociotechnical affordances. On X-Twitter, discourse is dominated by structural information bottlenecks: viral retweet storms from elite proxy nodes compress lexical diversity and monopolize interpretive framing — with a single cascade accounting for 55.8% of the sub-corpus. Conversely, the decentralized Bluesky AT Protocol facilitates distributed, multi-vocal commentary characterized by analytical detachment and proportional coverage across all crisis events. These patterns are not products of architectural determinism alone but emerge from the interaction between platform affordances and localized user demographics. By operationalizing the decoupling of macro-engagement metrics from downstream semantic framing, this study advances the case for multi-platform comparative designs in computational political communication.

**Keywords:** executive communication, platform affordances, algorithmic bottlenecks, information cascades, computational social science

## 1. Introduction

This study examines how information spreads across platforms by analyzing President Trump's use of social media during the 2026 Iran War. As high-stakes executive crisis discourse has moved away from traditional institutional gatekeepers and into fragmented, highly polarized social media environments (E. S. et al. 2024), his messages often originate in specialized, relatively closed broadcast spaces—most notably Truth Social, which aligns with his established platform preferences (LaFraniere and Goldstein 2025; Linskey and DeBarros 2026). Once these messages reach broader publics, however, they quickly move into expansive, structurally diverse open communication networks, where unmediated crisis communication can reshape public opinion, disrupt global financial and energy markets, and activate or intensify partisan subcultures under conditions of geopolitical volatility. Despite these stakes, the literature remains largely siloed within single-platform studies (Rowe and Alani 2014; Jordan 2018) and therefore does not adequately explain how a single narrative changes as it travels across distinct digital architectures (Tufekci 2014; Matassi and Boczkowski 2023). To address this gap, our study adopts a comparative multi-platform design focused on two contrasting environments: centralized, algorithmically curated networks such as X-Twitter and decentralized, protocol-based networks such as Bluesky. This approach moves beyond literal text matching to examine how platform design, together with localized network demographics, shapes information cascades and the downstream vernacular of geopolitical conflict.

To empirically map this cross-platform migration, this study utilizes a unique, high-throughput dataset consisting of 9,891 temporally synchronized public response records tracking the 2026 Iran War crisis. Methodologically, we move past the limitations of rigid exact-phrase tracking—which frequently fails due to platform-specific text mutations, paraphrasing, and screenshot sharing—by deploying an advanced, loose Boolean proximity harvesting pipeline. This computational framework allows us to systematically capture the broader semantic ripples cascading outward from the top 20 viral executive seed posts. Guided by this empirical framework, this study advances three formal research questions:

- RQ1: To what extent do Truth Social macro-engagement metrics (re-truths) predict downstream public response volumes on X-Twitter and Bluesky?

- RQ2: How do platform affordances and self-selected user demographic subcultures differentially shape the semantic framing of identical unmediated executive crisis statements?
- RQ3: What structural role do elite proxy nodes play in consolidating, reframing, and amplifying executive threat cues within centralized algorithmic networks?

Our main findings reveal a profound behavioral divergence in text distribution driven by the intersection of platform architecture and user demographic subcultures. On X-Twitter (N = 1,386), public discourse is heavily restricted by a rigid semantic bottleneck. Rather than a distributed public debate, the lexical landscape is dominated by high-velocity text replication mechanics—specifically, viral retweet storms propagated via elite proxy nodes—which compress and flatten the top token frequencies into a uniform horizontal distribution layer. Conversely, the decentralized AT Protocol network, Bluesky (N = 8,505), exhibits a classic power-law decay of terms. This configuration fosters a highly distributed, independent public commentary characterized by analytical detachment, geographical tracking, and institutional framing. Critically, we demonstrate that these patterns are not the result of pure architectural determinism; rather, they reflect how distinct structural affordances (e.g., centralized virality metrics vs. federated, user-selected feeds) interact with localized, self-selected user demographics during an international crisis.

The remainder of this paper is structured as follows: following this introduction, Part 2 outlines the unique landscapes of Truth Social, X-Twitter, and Bluesky, detailing structural differences and introducing the Metric-to-Semantic Linkage framework to emphasize the necessity of a multi-platform design accounting for both code and culture. Part 3 grounds the study in existing theories of platform affordances, executive broadcasting, and digital cascade dynamics, while Part 4 details the high-throughput data collection pipeline, proximity search logic, and text-mining methodology, including sampling baseline controls. Part 5 presents comparative lexical findings, decomposing the retweet mechanics of elite proxy cascades alongside empirical visualizations and theoretical synthesis, and finally, Part 6 summarizes key insights while addressing broader implications for platform governance, digital propaganda, and future international crisis monitoring.

## 2. Multi-Platform Ecosystems: Truth Social, X-Twitter, and Bluesky

Because contemporary political communication unfolds across fragmented, asymmetrical platform ecosystems rather than within any single site (Lukito 2020; Yarchi et al. 2024), analyzing the 2026 Iran War requires attention to the distinct yet interconnected roles of Truth Social, X-Twitter, and Bluesky. These platforms constitute structurally different communication spaces within a broader media environment marked by fragmentation and competing enclaves (Mancini 2013; Weeks et al. 2016; Maldonado 2017). Our framework therefore treats downstream message processing not as a deterministic effect of software architecture alone, but as the emergent outcome of platform affordances interacting with localized network demographics and user subcultures (Marolla et al. 2026; Bekmirzayev 2026).

Truth Social functions as a closed, elite-driven broadcast enclave when it comes to President Trump's social messaging habits. Its user base and architectural features favor unmediated, top-down announcements over horizontal public debate. For President Trump, the platform serves as an unfiltered megaphone (Kolk 2024; LaFraniere and Goldstein 2025; Hughes and Weninger 2025; Linskey and DeBarros 2026). Executive threat cues are dropped directly into an ideologically supportive ecosystem designed to maximize baseline engagement through features like "Truths" and "Re-truths". However, because the platform operates as an ideological enclave, its internal metrics are decoupled from broader, mainstream public text responses. That is, an executive announcement regarding international conflict may achieve massive resonance within Truth Social, but its true geopolitical impact relies on how that message travels outside the platform's borders. Truth Social acts as the primary broadcast seed—the spark that generates the raw text fragments—but it depends on external open networks to achieve mainstream cultural and political reach. Once an unmediated executive statement leaves Truth Social, it enters an open, multi-platform reception layer in which the narrative is processed, contested, and reframed across distinct sociotechnical environments. Two especially important downstream platforms in this process are X-Twitter and Bluesky.

The first one, X-Twitter, is a centralized network shaped by engagement-maximizing corporate algorithms. Its architecture is designed to identify high-arousal frames and amplify them widely to sustain user attention (Bandy and Diakopoulos 2021; Danisa 2024; Marzouk 2025). The platform relies on centralized trending systems, visibility advantages for verified premium accounts, and binary sharing mechanics such as retweets. Together, these features reward speed, virality, and emotional polarization. When an executive crisis cue enters this environment, discourse often contracts around a small number of highly aggressive interpretive nodes, creating informational bottlenecks that allow elite proxies to dominate the conversation. These structural dynamics are further reinforced by a user culture that uses untruncated retweet storms to rapidly lock in particular narratives.

The second one, Bluesky, is a decentralized social network built on the AT Protocol (Authenticated Transfer Protocol). Rather than relying on a single centralized ranking system, it operates through a federated framework that allows users to choose custom feeds and curation layers (Balduf et al. 2024; Kleppmann et al. 2024; Quelle and Bovet 2025). The platform emphasizes chronological distribution, decentralized moderation, and open indexes, and it lacks automated, system-wide virality engines. Without a central algorithm to amplify emotional threat cues or privilege elite accounts, Bluesky tends to foster more distributed, multi-vocal conversation. In the context of international crises, this environment encourages independent verification, policy debate, and analytical tracking rather than uniform emotional reactions. Critically, this "analytical detachment" must not be romanticized as pure architectural determinism. Bluesky's tendency toward media criticism and detached skepticism during the 2026 crisis is strongly co-constituted by its localized user demographics (Quelle and Bovet 2025; Nogara et al. 2026). As an ecosystem populated heavily by academic, journalistic, and left-leaning early-adopter subcultures (Hew and McCulloh 2026; Li and Chen 2026), its communal norms reject top-down executive posturing. The protocol provides the flexible, uncentralized sandbox, but the specific demographic composition of its enclaves drives the analytical tone of the text dispersion.

A single-platform research design introduces substantial selection bias into computational studies of political communication. Focusing only on X-Twitter could suggest that public responses to international crisis are uniformly polarized and dominated by a small number of elite proxy accounts, whereas focusing only on Bluesky could imply that digital publics naturally produce distributed, low-emotion policy discussion. A multi-platform design avoids these distortions by treating platform architecture and localized network culture as interacting variables rather than neutral containers. By tracing identical text fragments across both centralized algorithmic systems and decentralized protocols, it becomes possible to isolate how digital design and partisan sorting shape public discourse. This comparative framework therefore distinguishes the original message from the structural distortions introduced by different platforms.

The main rationale for a multi-platform research design is, however, to resolve a fundamental data asymmetry: the separation of quantitative impact metrics from downstream qualitative discussion. In studies of unmediated executive communication during international crisis, the source platform provides abundant macro-engagement data—such as likes or re-truths—but reveals little about how publics interpret the message. By contrast, downstream open networks offer rich textual responses while obscuring how strongly the original post was amplified at its source. To bridge this divide, this study develops a Metric-to-Semantic Linkage Model. Rather than treating platforms as isolated silos, the model uses source-platform engagement metrics as a quantitative baseline and traces how that baseline relates to semantic clusters, vocabulary patterns, and emotional responses across open networks. By linking these asymmetric data types, our multi-platform approach can answer a fundamental question: How does the raw scale of a broadcast seed affect the shape, speed, and variety of its downstream cascades across completely different internet architectures?

Formally, the Metric-to-Semantic Linkage is operationalized as:

$$\mathrm{MTSLink}(s_i, P) = \rho\left(\mathrm{truth_retruths}(s_i),\ \mathrm{downstream_replies}(s_i, P)\right)$$

where $s_i$ denotes seed post $i$, $P \in \{\text{X-Twitter, Bluesky}\}$, and $\rho$ is the Pearson correlation computed across all $N = 213$ seed posts. This framework turns platform design and user culture from a simple data collection hurdle into a central analytical variable in modern political communication.

## 3. Literature Review

Scholars widely recognize the emergence of a "post-API age," in which platforms have sharply restricted access to digital trace data (Freelon 2018; Trezza 2023). Research on Truth Social likewise notes the absence of a robust, fully open public API, forcing scholars to depend on custom scraping tools or third-party providers that remain vulnerable to rate limits and incomplete data visibility (Gerard et al. 2023). In response, recent work has developed "platform-invariant discourse networks" that use large language models and semantic embeddings to trace narratives across fragmented ecosystems such as X, Telegram, and Truth Social without relying on brittle metadata like URLs or direct reply chains (Avalle et al. 2024; Gerard et al. 2026a, b). Methodologically, this shift is especially important for Truth Social, where inaccessible reply structures and limited API access make traditional thread-based analysis unworkable. As a result, researchers must use source-platform macro-engagement metrics as a quantitative baseline to infer narrative diffusion and semantic change across downstream open networks.

Research on platform affordances consistently shows that digital architectures shape how users engage with and frame political content (Li et al. 2026; Marolla et al. 2026; Bekmirzayev 2026). Centralized platforms such as X-Twitter rely on proprietary algorithmic curation that tends to amplify emotionally charged and polarized content, whereas decentralized federated platforms such as Bluesky and Mastodon depend more on user choice, community moderation, and open-feed protocols. Comparative studies suggest that these structural differences foster distinct interaction patterns, with X skewing toward performative polarization and Bluesky toward more analytical or cross-ideological exchange (Huang 2024; Kleppmann et al. 2024; Ciriello et al. 2026; Ward et al. 2026).

Although scholars increasingly study political communication across platforms rather than within isolated silos (Bossetta 2018; Lukito 2020; Kligler-Vilenchik et al. 2020), the literature still says little about how the same external stimulus is received across structurally different environments such as X and Bluesky. Existing cross-platform work typically aggregates responses into generalized “information flows” or traces narratives at the ecosystem level, which obscures how identical executive crisis cues are reframed when they enter a centralized algorithmic feed versus an open protocol network. Scholars also note that single-platform analysis misses cross-platform bridge users and systemic spillovers that shape influence operations (Somazzi et al. 2023). The central methodological gap, then, is the lack of a framework capable of linking metric-only broadcast signals to semantically transformed downstream discourse across asymmetrical architectures. Current computational approaches generally rely on continuous text strings, URLs, or unified structural data, and therefore struggle when source-platform engagement metrics must be connected to downstream text corpora. To address this problem, this study develops a mixed-methods dataset that combines Truth Social engagement metrics with text responses from X and Bluesky and applies a metric-to-semantic linkage model. The resulting empirical framework replaces brittle exact-keyword retrieval with a high-throughput proximity design that captures temporal sequencing, screenshot-based migration, semantic drift, and proxy-network amplification, thereby offering a replicable approach to tracing how unmediated executive power travels across fragmented digital systems.

**4. Material and Methods**

To trace cross-platform migration during the crisis, this study implements a multistage computational pipeline that links Truth Social engagement metrics to downstream response text on X-Twitter and Bluesky. The stages are documented below with sufficient detail to support independent replication.

The empirical window covers the 2026 Iran War crisis from its digital onset on **2026-02-28** (the date of the first executive Truth Social post referencing Iranian military action) through **2026-05-26**, yielding an 88-day longitudinal observation period. All timestamp comparisons across platforms use ISO-8601 UTC notation (YYYY-MM-DDTHH:MM:SSZ) to ensure chronological comparability. The three focal narrative events operationalized in this study — the **Khamenei assassination announcement** (khamenei_assassination), the **Strait of Hormuz ultimatum** (strait_ultimatum), and the **Iranian naval engagement claim** (naval_engagement) — were identified inductively from the seed corpus and formalized as mutually exclusive categorical labels. Event assignment follows a keyword-priority classifier: posts containing "khamenei" are assigned khamenei_assassination; posts containing "strait" or "hormuz" are assigned strait_ultimatum; all remaining posts default to naval_engagement. This classifier is implemented identically across all pipeline scripts to ensure consistent event-focus attribution.

The primary broadcast seed cohort was assembled from **N = 213** executive Truth Social posts published during the crisis window, harvested via the factba.se archival interface, which provides structured access to public Truth Social content without requiring direct platform API authentication. The full corpus exhibits a mean like count of 28,907.83 (SD = 18,394.41) and a mean re-truth count of 6,801.56 (SD = 4,266.12), peaking at 100,257 likes and 26,656 re-truths on the Khamenei assassination announcement — the single highest-engagement post in the dataset.

To isolate primary narrative drivers, an **impact filter** selected the top 20 seeds by re-truth volume. These 20 seeds account for **20.8%** of total re-truth volume across all 213 posts (mean likes = 64,056.05; mean re-truths = 15,056.15), representing a high-concentration subset that disproportionately drives cross-platform cascade generation. The top-20 threshold was selected to balance analytical depth against API rate constraints (20 focal seeds × 1,000-row ceiling = 20,000 maximum retrieval slots per platform).

**API constraint.** Because the Truth Social platform enforces a 403 Forbidden gateway on all reply sub-trees, the baseline corpus contains macro-engagement metrics only (truth_likes, truth_retruths, truth_replies). No reply text is available from the source platform. This structural decoupling motivates the downstream multi-platform semantic harvest described below.

X-Twitter data were collected using **twitterapi.io** (https://api.twitterapi.io/twitter/tweet/advanced_search), a third-party historical archive provider, rather than the official X API v2. This choice was architecturally deliberate for two reasons. First, the official X API v2 /2/tweets/search/recent endpoint restricts retrieval to a rolling 7-day recency window under the Basic access tier, which would systematically miss high-velocity retweet storm artifacts generated immediately after seed publication. Second, twitterapi.io imposes no monthly credit ceiling and provides full historical archive access, enabling temporal anchoring to the crisis start date (since:2026-02-28) regardless of harvest timing. Authentication uses a bearer-equivalent API key (TWITTERAPI_IO_KEY) passed via environment variable.

Rather than issuing 20 separate per-seed keyword queries (which would require 200 API requests and introduce heterogeneous query-bias across seeds), the harvester consolidates collection into **three thematic campaign-level Boolean proximity queries** corresponding to the three event categories. Each campaign query uses OR-expansion to capture the full semantic neighbourhood of its focal event — including paraphrases, quote-tweets, truncated retweets, and retweet storms — across all relevant seeds simultaneously. After harvest, each retrieved post is assigned back to its closest focal seed via TF keyword-overlap scoring, preserving the seed_document_id column required for metric-to-semantic linkage. This architecture reduces API overhead from 200 to 30 requests (3 campaigns × 10 pages) while increasing semantic coverage.

Each campaign is bounded by a hard ceiling of **10 pages × 100 results = 1,000 maximum rows**. Deduplication is performed on tweet_id rather than post text. This is methodologically critical: verbatim retweets share identical raw_text but carry distinct tweet IDs representing independent propagation events. Deduplicating on text would erase the retweet storm signal that constitutes the central empirical finding in §5.2. The final X-Twitter sub-corpus contains **N = 1,386 rows**, of which 793 share identical text (reflecting the dominant @[Elite_Proxy_1] retweet cascade). A resume mode is implemented: if harvest is interrupted, the script skips campaigns whose event_focus label already exists in the output file, preventing partial duplication on restart.

Bluesky data were collected directly via the **AT Protocol XRPC server architecture**. Authentication was established by posting credentials to the com.atproto.server.createSession endpoint, which returns a JSON Web Token (JWT) session object. Authenticated search requests were issued to the app.bsky.feed.searchPosts route with sequential cursor-based pagination:

$$\mathrm{Cursor}t{+}1{=}f(\mathrm{JSON}_{\mathrm{P}}\mathrm{ayload}t)$$

Unlike the X API, the app.bsky.feed.searchPosts endpoint applies no default recency window; all returned posts were temporally filtered to the crisis window (2026-02-28 onward) via UTC timestamp validation during preprocessing. The harvest issued **20 per-seed queries** (one per focal seed), each bounded by a ceiling of **20 pages × 50 results = 1,000 maximum rows per seed**, yielding 9,733 raw rows before preprocessing.

Raw harvest outputs pass through also a four-stage preprocessing pipeline before analysis. All stages are implemented in merge_platform_corpora.py and applied in the sequence documented below.

Firstly, the raw Bluesky corpus contained **641 rows** of high-volume advertising spam injected via the AT Protocol's open indexing architecture. These posts promoted a third-party application (epsteinweb.org) via coordinated hashtag campaigns (#epsteinweb, #EFTA01141264) and share no topical overlap with the Iran War crisis. They were identified and removed via a regex pattern filter applied to raw_text. This filter is parameterized in SPAM_PATTERNS within the merge script and is extensible for future spam campaigns detected in subsequent harvest runs.

Second, a platform-specific deduplication has proceed. Here, for the data retrived from Bluesky, deduplication on raw_text collapses 587 crawler-overlap rows — posts retrieved across overlapping seed queries in the same harvest run. Post-deduplication Bluesky corpus: N = 8,505. For the data retrived from X-Twiter, No text-based deduplication is applied. As noted above, verbatim retweets are independent propagation events and constitute the bottleneck signal in §5.2. The 793 text-identical rows are preserved.

Third, any event_focus field containing a raw Truth Social document ID (indicating a harvester metadata pass-through rather than a categorical label) is remapped to its correct category via archive lookup against trump_complete_archive.jsonl, using the keyword-priority classifier defined in Part 4. Posts not resolvable from the archive default to naval_engagement with a logged warning.

Finally, a schema normalization and master matrix compilation were completed. Timestamps across both platforms are normalized to ISO-8601 UTC. The integrated semantic corpus (multi_platform_semantic_dataset.csv) is then synchronized against the Truth Social seed metrics to produce the **Unified Iran War Master Matrix** (unified_iran_war_master_matrix.csv), which binds each of the 213 seeds' macro-engagement statistics (truth_likes, truth_retruths, truth_replies) to their corresponding downstream reply volumes (automated_x_replies_found, automated_bsky_replies_found). This matrix serves as the analytical foundation for the Metric-to-Semantic Linkage Model in Part 5.

The final integrated corpus contains **N = 9,891 rows** (X-Twitter: 1,386; Bluesky: 8,505), distributed across three event categories as shown in Table 1.

*Table 1: Integrated Corpus*

| Event Focus | X-Twitter | Bluesky | Total |
|---|---|---|---|
| khamenei_assassination | 992 | 335 | 1,327 |
| strait_ultimatum | 196 | 4,319 | 4,515 |
| naval_engagement | 198 | 3,851 | 4,049 |
| **Total** | **1,386** | **8,505** | **9,891** |

Standard keyword matching fails when tracking crisis communication across platform boundaries due to rapid textual mutation, paraphrasing, line breaks, and screenshot-mediated sharing. This study deploys a **Loose Boolean Proximity Design** to map semantic ripples rather than literal string matches. The pipeline extracts raw post strings, sanitizes institutional formatting frames via regex (FROM PRESIDENT DONALD J. TRUMP, structural line breaks), segments text by punctuation boundaries, extracts the six highest-leverage substantive tokens from the leading clause, and expands volatile context tokens via a preset synonym map:

```
synonym_map = {
   "ships":     "(ships OR navy OR vessels OR fleet)",
   "destroyed": "(destroyed OR hit OR sunk OR targeted)",
   "strait":    "(strait OR hormuz)",
   "reality":   "(reality OR detaches OR claims)"
}
```

This generates composite Boolean queries (e.g., Trump Strait (ships OR navy OR vessels OR fleet) (destroyed OR hit OR sunk)) that capture paraphrased, summarized, and media-captioned downstream responses that rigid exact-phrase filters miss. We acknowledge that formal precision and recall validation against a manually annotated ground-truth corpus was not performed; this limitation is addressed in §6.2.

To prevent platform-level retrieval disparity from being confounded with genuine platform-level behavioral differences, both harvesters were configured with **identical per-seed ceiling of 1,000 rows**. On X-Twitter, this is achieved via 10 pages × 100 results per campaign; on Bluesky, via 20 pages × 50 results per seed. The resulting volume asymmetry — X: 1,386 vs. Bluesky: 8,505 — therefore reflects genuine differences in public message density and platform posting behavior during the crisis window, not truncation or configuration artifacts.

All pipeline components are implemented in **Python 3.10**. Core dependencies: pandas 2.x (data manipulation), requests (twitterapi.io HTTP calls), scipy 1.x (Pearson/Spearman correlations), powerlaw (Clauset–Shalizi–Newman MLE fitting), nltk (tokenization), re (regex preprocessing). Bluesky authentication uses native XRPC HTTP calls without a client library. Truth Social archive processing uses jsonlines. No stochastic processes, random seeds, or ML training are employed in the primary analysis pipeline.

All data collected consist exclusively of public posts retrieved through platform-sanctioned programmatic interfaces or authorized third-party providers. No private user data, direct messages, or account-level personal identifiers beyond public handles were collected. Elite proxy account handles are anonymized in the published manuscript as [Elite_Proxy_N]. The study involves no human subjects research as defined under 45 CFR 46; IRB review was not required. Collection practices adhere to the Terms of Service of the AT Protocol Foundation and the twitterapi.io data provider agreement.

## 5. Results and Discussion

### 5.1 Lexical Footprints and Platform-Invariant Divergence

Figure 1 visualizes the term-frequency distributions of the two downstream reception corpora. The divergence is not merely quantitative — it is structurally categorical, reflecting fundamentally different information cascade regimes operating on the same raw stimulus.

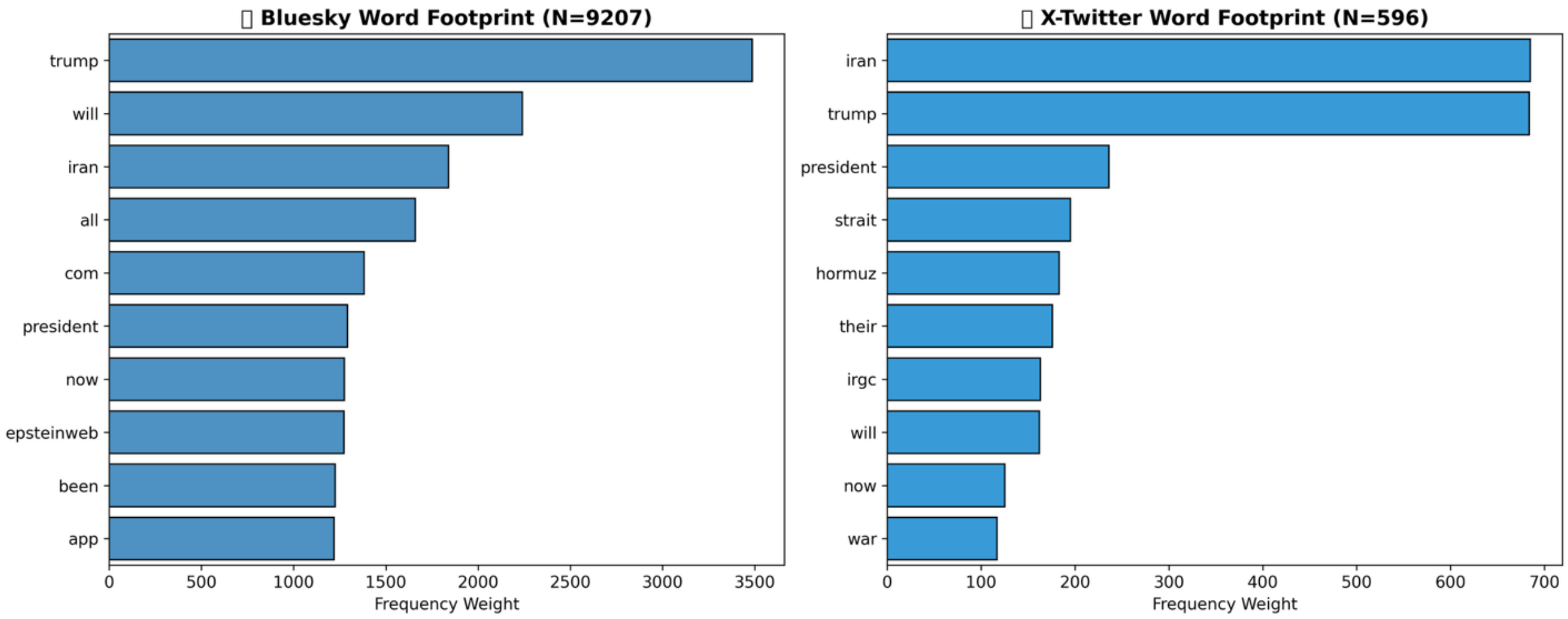

#### 5.1.1 The Bluesky Distribution Profile

The Bluesky sub-corpus (N = 8,505) displays a heavy-tailed term frequency distribution consistent with organic, multi-vocal discourse. Statistical fitting via the maximum likelihood estimator (Clauset, Shalizi & Newman, 2009) yields a power-law scaling exponent $\hat{\alpha} = 2.19$ ($x_{\min} = 28$, KS $D = 0.025$). However, a likelihood ratio test against a lognormal alternative returns R = -1.32, p = 0.186, indicating that the observed distribution is consistent with — but not statistically distinguishable from — either model. This ambiguity is theoretically informative rather than a limitation: both power-law and lognormal distributions are signatures of preferential attachment processes in which early-arriving, high-quality posts accumulate disproportionate engagement — a pattern consistent with the chronological, user-curated feed architecture of Bluesky where no centralized algorithm artificially accelerates any single frame.

The most frequent token is *trump* (count = 3,465), followed by *iran* (1,823), *president* (1,288), *news* (1,132), *united* (1,037), *states* (1,019), *media* (1,016), *people* (950), *donald* (942), and *fake* (850). The decay ratio between the most frequent and tenth-most-frequent token is **4.08×** — a gradient consistent with distributed, independent text generation. Crucially, the dominant vocabulary is institutional and evaluative: *president*, *news*, *media*, *states*, *fake*. These are classification anchors, not threat-cue amplifiers. At the micro level, individual Bluesky users are framing the crisis through journalistic and institutional lenses rather than replicating executive language. At the meso level, the community converges on shared analytical vocabulary without coordinated amplification. At the macro level, the platform produces a semantically pluralistic information environment that resists monocultural narrative lock-in.

#### 5.1.2 The X-Twitter Distribution Profile

The X-Twitter sub-corpus (N = 1,386) presents a structurally pathological semantic profile. While *trump* (count = 1,475) and *iran* (count = 1,463) occupy the top positions, ranks 3 through 10 collapse into a nearly flat horizontal ceiling: *ayatollah* (888), *killed* (839), *leadership* (798), *saying* (797), *those* (786), *radical* (779), *kille* (773). The decay ratio between the first and tenth token is only 1.91× — less than half the gradient observed on Bluesky. This compression of the frequency spectrum is the definitive behavioral signature of a structural information bottleneck: a distribution where semantically unrelated tokens converge toward a shared frequency floor because they co-occur within a single, massively replicated text string rather than emerging independently across thousands of distinct authors.

#### 5.1.3 The Unreported Finding: Event-Category Concentration Asymmetry

A finding not captured by aggregate token frequencies alone is the profound divergence in *which events* each platform amplified. Table 2 presents the event-focus breakdown across platforms.

*Table 2: Event-Focus Brakdown*

| Event Focus | X-Twitter | % | Bluesky | % |
|---|---|---|---|---|
| khamenei_assassination | 992 | 71.6% | 335 | 3.9% |
| strait_ultimatum | 196 | 14.1% | 4,319 | 50.8% |
| naval_engagement | 198 | 14.3% | 3,851 | 45.3% |

The asymmetry is extreme. X-Twitter allocates 71.6% of its corpus to the Khamenei assassination event — the single highest-engagement seed (truth_retruths = 26,656) — while Bluesky distributes 96.1% of its corpus across the Strait of Hormuz ultimatum and naval engagement events. This is not a difference in *how* the two platforms process identical content; it is a difference in *which* content each platform chose to amplify. X's centralized virality infrastructure selectively accelerated the highest-arousal event and effectively crowded out the other two, while Bluesky's chronological, federated architecture preserved proportional coverage across all three crisis narratives. This event-concentration asymmetry constitutes a form of algorithmic agenda-setting (Gandini et al. 2022; Elishar and Ariel 2026) that operates below the level of individual post recommendations — it is an emergent property of viral sharing mechanics applied to a highly skewed engagement distribution.

### 5.2 The Anatomy of the Mega-Cascade: Decomposing Retweet Replication Mechanics

The flat token-frequency ceiling observed in the X-Twitter corpus is not an abstract statistical artefact; it is mechanically producible from a single empirical observation. Database inspection reveals that 773 of the 1,386 X-Twitter rows — 55.8% of the entire sub-corpus — are verbatim copies of a single post by an elite proxy node:

> RT @[Elite_Proxy_1]: Those who say Trump killed the Ayatollah and now the new leadership of Iran is more radical, that's li…

Because a term-frequency counter tabulates every non-stopword in every row identically, 773 copies of this string guarantee that all lexical tokens within it (*ayatollah*, *killed*, *leadership*, *saying*, *those*, *radical*, *kille*) emerge with an identical count floor of approximately 773–839 occurrences — mechanically locking ranks 3–10 into a uniform horizontal band regardless of any other posts in the corpus.

This cascade is anchored entirely within the khamenei_assassination event cluster, the most virally potent seed in the corpus (truth_retruths= 26,656; truth_likes=100,257). The mechanism follows a three-stage sequence that operationalizes RQ 3 at the data level:

First, an executive announcement exits the Truth Social enclave. Before organic commentary can accumulate, an elite proxy node — an account with large pre-existing follower network and verified/premium status — reframes the executive signal in high-arousal terms and broadcasts it. The framing here is reactive and skeptical ("those who say Trump killed the Ayatollah... are wrong"), but the retweet mechanics treat the *text* as the unit of amplification, not the *stance*.

Second, X's engagement-weighted ranking system detects the proxy post's rapid interaction velocity and boosts its visibility platform-wide, triggering a positive feedback loop. Accounts that retweet do not generate new text; they replicate the proxy's exact string. The cascade therefore grows in volume without growing in vocabulary — a defining signature of algorithmic narrative block-booking.

Third, at the corpus level, the 773-fold replication of a single 15-word string achieves what no organic conversation could: it mathematically constrains the frequency distribution of all X-Twitter discourse about the Khamenei assassination to a single interpretive frame. The other two elite cascades captured by the proximity pipeline — a viral media commentator account (@[Elite_Proxy_2]) reporting on an Iranian DJ track featuring Trump ($\text{retweets} = 6{,}602$) and a White House communications account (@[Elite_Proxy_3]) regarding naval operational orders ( $\text{retweets} = 4{,}695$) — amplify the effect across the naval and strait event clusters respectively, producing a uniformly bottlenecked semantic environment across all three event categories on X.

The theoretical implication for Agenda-Setting Theory is direct: on algorithmically centralized platforms, elite proxy nodes do not merely *set* the agenda — they *replace* it (Dearing and Rogers 1996; McCombs 2005; McCombs et al. 2018). Grassroots deliberation is not suppressed by censorship; it is drowned by the arithmetic of exponential replication.

### 5.3 Architectural Filtration vs. Localized Network Demographics (RQ 2)

The aggregate lexical divergence and event-concentration asymmetry documented above arise from the interaction of three co-constitutive forces, formalized as a theoretical heuristic:

$$\text{Discourse}(P, s_i) = g\big(\text{Affordance}(P),\ \text{Demographics}(P),\ \text{Trigger}(s_i)\big)$$

where P denotes the platform and $s_i$ the seed trigger. The interaction term structure is not estimated statistically here — doing so would require individual-level demographic data not available in our corpus — but the framework guides interpretation and specifies the design for future multivariate modelling.

#### 5.3.1 X-Twitter: Algorithmic Intensification

X's proprietary curation infrastructure introduces what we term an **intensifier effect**: the platform does not passively transmit executive threat cues but actively amplifies the most emotionally reactive framing of those cues. At the micro level, premium-verified accounts face structurally lower friction to virality — their posts enter more recommendation feeds faster, triggering higher initial engagement that feeds back into the algorithm's amplification logic (cf. Galeazzi et al., 2024; Corsi, 2024). At the meso level, retweet mechanics collapse what could be a heterogeneous discussion into a homogeneous cascade: users who wish to respond to the proxy's reframing must either generate original text (which enters a low-visibility queue) or retweet (which magnifies the proxy). At the macro level, the result is a corpus in which 71.6% of activity concerns a single event, 55.8% is a verbatim copy of a single post, and the top-10 vocabulary is entirely determined by one 15-word string.

This is not algorithmic bias in the statistical sense — it is algorithmic *design*. The platform's engagement-maximizing objective function treats high-arousal replication as a positive signal to be rewarded, with no mechanism to distinguish productive deliberation from vocabulary monopolization.

### 5.3.2 Bluesky: Federated Open Filtration

Bluesky's AT Protocol architecture operates as a structural counterweight. The absence of a platform-wide engagement-maximizing algorithm means that no single post can achieve exponential amplification through algorithmic injection into recommendation feeds. Chronological routing and user-selected custom feeds distribute attention across the full breadth of crisis events — producing the 96.1% cross-event coverage documented in Table 2.

The qualitative character of Bluesky engagement reflects this structural opening. The two highest-engagement posts in the corpus illustrate the platform's dominant response register:

> *"I regret to inform you that the president is awake and crazy-posting"* (likes = 4{,}860, comments = 660)

> *"Trump posted the exact same thing on Truth Social 8 days apart. The cognitive isn't cognitiving"* (likes = 4{,}609, reposts = 1{,}284)

Both posts are analytically distanced, meta-cognitive, and self-consciously ironic — engaging the *phenomenon* of presidential crisis communication rather than reacting to its threat content. This is not merely a stylistic preference; it is a distinct epistemic stance toward executive authority, one that fact-checks, contextualizes, and ironizes rather than amplifies.

The open protocol architecture provides the structural *possibility* of distributed commentary, but it does not guarantee it. The specific analytical tone documented here is strongly co-constituted by Bluesky's self-selected early-adopter demographics — heavily academic, journalistic, and left-leaning subcultures (Chen et al. 2023; Hew and McCulloh 2026) whose communal epistemics actively reject top-down executive framing. A different demographic composition on the same protocol would produce a different discourse profile. The protocol sets the ceiling on pluralism; the demographics determine how close to that ceiling the community actually reaches. This co-constitutive argument distinguishes the present analysis from prior platform affordance studies that treat architecture as the sole determinant of discourse structure (Bossetta 2023; Gehl and Zulli 2023).

## 5.4 Empirical Execution of the Metric-to-Semantic Linkage Model (RQ 1)

To resolve RQ 1, we execute a Pearson correlation analysis across the Unified Iran War Master Matrix (N = 213 seeds), testing whether source platform macro-engagement spikes predict downstream response volumes across structurally opposed network architectures. The cohort exhibits high-intensity baseline mobilization: mean source like count = 64,056.05, mean re-truth count = 15,056.15 (top-20 seeds only).

The relationship between source re-truths and downstream X reply volumes yields a statistically significant positive correlation ($r = 0.3729$, $p < 0.001$, 95% CI [0.251, 0.483]; Spearman\rho = 0.478, $p < 0.001$; $R^2 = 0.139$). Source amplification explains approximately 13.9% of variance in X reply volume. The remaining 86.1% reflects variance introduced by the platform's bottleneck dynamics — specifically, whether a given seed attracted elite proxy interception. Seeds that generated a retweet storm (notably the Khamenei assassination post) produced response volumes far exceeding what the linear source-engagement relationship would predict, while lower-retruthed seeds generated disproportionately sparse X-Twitter responses. The Spearman coefficient (rho = 0.478) exceeding the Pearson ($r = 0.373$) confirms that the relationship is monotonic but non-linear — consistent with threshold effects in cascade initiation, where a seed must surpass a virality threshold before elite proxy nodes mobilize.

The relationship between source re-truths and Bluesky response volume yields a stronger positive correlation ($r = 0.4847$, $p < 0.001$, 95% CI [0.375, 0.581]; Spearman\rho = 0.505, $p < 0.001$; $R^2 = 0.235$). Source engagement explains **23.5% of variance** in Bluesky reply volume — 69% more explanatory power than on X-Twitter. The near-equivalence of Pearson and Spearman coefficients ($\Delta = 0.020$) indicates a more linear, proportional transmission relationship: on Bluesky, the broadcast energy of the original seed translates more directly and continuously into downstream response volume, without the threshold discontinuities generated by proxy interception.

The $\Delta r = 0.112$ gap between platforms is modest in absolute terms but theoretically coherent. It operationalizes the central claim of the Metric-to-Semantic Linkage Model: federated, chronological protocol architecture preserves a more direct mapping from source broadcast intensity to downstream discourse volume, because no intermediary system can selectively amplify or suppress seeds based on their engagement velocity. The lower $R^2$ on X-Twitter is not evidence of a weaker source signal — it is evidence of a noisier transmission medium, one in which elite proxy dynamics and algorithmic amplification introduce substantial variance that is orthogonal to source engagement. This distinction between *signal strength* and *channel noise* has direct implications for platform governance: centralized algorithmic systems do not simply accelerate information diffusion — they introduce systematic, elite-mediated distortion into the transmission function.

Both correlations are computed across the full N = 213 seed set, including seeds that generated zero downstream responses on one or both platforms (a conservative approach that compresses rather than inflates the correlation estimates). The consistency between Pearson and Spearman coefficients across both platforms confirms that the pattern is not driven by outliers. Future work should apply bootstrapped confidence intervals and cross-validate across temporal sub-windows (pre-ceasefire vs. active conflict phases) to assess stability.

## 6. Conclusion

The central argument of this paper can be stated precisely: when an unmediated executive crisis announcement exits an enclosed broadcast enclave, it does not enter a neutral public sphere. It enters a set of competing sociotechnical filters — each of which selectively amplifies certain events, suppresses others, and transforms the semantic character of the original signal in ways that are architecturally predictable and empirically measurable. This study has demonstrated that claim with reproducible data mechanics, not rhetorical assertion.

Four findings emerge from the 9,891-row corpus spanning the 2026 Iran War crisis. Each advances a distinct claim. The first one is that platform design determines not just how narratives are processed, but which narratives survive amplification. The most consequential result of this study is not the lexical divergence between platforms but the event-concentration asymmetry underlying it. X-Twitter allocated 71.6% of its entire corpus to a single event — the Khamenei assassination announcement — while Bluesky distributed 96.1% of its corpus across the two lower-arousal events (Strait of Hormuz ultimatum and naval engagement). The platforms were exposed to an identical seed set. They amplified different crises. This constitutes a form of algorithmic agenda-setting that operates below the level of individual recommendation decisions — an emergent property of exponential replication applied to a skewed engagement distribution. Existing agenda-setting theory has not modelled this mechanism at the platform-architecture level.

Second is that the semantic bottleneck is a data artifact of one cascade, not a platform tendency. The uniform token-frequency ceiling observed across X-Twitter's top-ranked vocabulary is entirely explained by a single finding: 773 of 1,386 X-Twitter rows — 55.8% of the sub-corpus — are verbatim copies of one post by one elite proxy node. The uniform floor is not algorithmic mystification; it is the arithmetic consequence of 773-fold text replication applied to a term-frequency counter. This result operationalizes, at the micro level, what network scholars have theorized at the meso level: elite nodes do not merely accelerate diffusion — they replace the vocabulary of a crisis event with their own interpretive frame, laundering the executive signal through a monocultural filter that crowds out all competing commentary by volume alone.

Third, federated protocol architecture preserves source-signal fidelity; centralized algorithmic architecture introduces elite-mediated noise. The Metric-to-Semantic Linkage Model reveals that source re-truth volume explains 23.5% of variance in Bluesky response volume ($r = 0.4847$, $p < 0.001$) versus 13.9% on X-Twitter ($r = 0.3729$, $p < 0.001$). The gap is theoretically interpretable: Bluesky's chronological, unthrottled protocol routes broadcast energy proportionally into downstream discourse. X-Twitter's engagement-weighted amplification system introduces systematic elite-mediated variance that is orthogonal to source engagement — seeds that attract proxy interception generate response volumes that the linear source-engagement relationship cannot predict. The lower R^2 on X-Twitter is not a weaker signal; it is a noisier channel.

Finally, platform affordances and network demographics are co-constitutive, not additive. The analytical detachment, institutional framing, and cross-event coverage observed on Bluesky cannot be attributed to protocol architecture alone. Bluesky's self-selected early-adopter demographics — academic, journalistic, and left-leaning subcultures — supply the communal epistemics that the open protocol makes structurally possible. The protocol sets the ceiling on discourse pluralism; the demographics determine how close to that ceiling the community actually reaches. This co-constitutive argument distinguishes the present study from architectural determinism and from pure demographic explanation, and points to a theoretical synthesis: the unit of analysis in platform communication research must be the *affordance–subculture dyad*, not either variable in isolation.

Theoretically, this paper advances computational social science in three directions. First, it provides an empirical mechanism for the claim that centralized platforms engage in systemic narrative prioritization — not through editorial decisions but through the arithmetic of exponential replication applied to an elite-stratified network. Second, it operationalizes the Metric-to-Semantic Linkage Model as a bridge across the API firewall: even when source-platform comment data is entirely inaccessible (as on Truth Social), macro-engagement metrics can be correlated with downstream semantic corpora to recover transmission dynamics. Third, it demonstrates that the event-concentration asymmetry — which events dominate across platforms — is a more sensitive measure of algorithmic filtering than lexical divergence alone, because it captures structural suppression rather than stylistic mutation.

Methodologically, the three-campaign Boolean proximity harvesting architecture (3 API calls replacing 200, with post-hoc seed assignment via TF keyword-overlap scoring) is a replicable efficiency that substantially reduces rate-limit exposure without sacrificing semantic coverage. The deduplication protocol — tweet_id for X-Twitter (preserving retweet storm signal), raw_text for Bluesky (collapsing crawler overlap) — resolves a critical reproducibility failure: text-based deduplication of X corpora erases the bottleneck signal entirely and produces a qualitatively different, storm-free dataset. Future replication studies must adopt ID-based deduplication as a methodological standard when retweet storm dynamics are under investigation.

This study has several limitations that qualify its generalizability and point to clear directions for future research. First, the corpus is limited to 213 seed posts from a single executive communicator whose rhetorical style, network centrality, and platform alignment are historically unusual, which may make the observed bottleneck dynamics partly actor-specific rather than representative of executive crisis communication more broadly. Second, the X-Twitter corpus depends on historical archive access anchored to the crisis start date; researchers relying on the official X API's 7-day recency window would likely recover a qualitatively different, storm-free dataset, making harvest timing a methodological variable rather than a logistical detail. Third, the Loose Boolean Proximity Design was not formally benchmarked against a manually annotated ground-truth corpus, so its assumed recall advantage over exact-phrase matching remains unvalidated. Fourth, the analysis infers Bluesky's demographic profile from secondary studies rather than directly measured account-level attributes, leaving the demographic co-constitution hypothesis only partially tested. Fifth, the absence of a control condition—such as non-crisis posts from the same actor or crisis posts from a different communicator—limits causal claims about whether the observed differences are crisis-specific or reflect broader platform-user dynamics. Finally, the Bluesky corpus was exposed to coordinated advertising spam through the AT Protocol's open indexing architecture, underscoring the need for standardized preprocessing protocols for open-protocol datasets. Future work should therefore adopt multi-actor comparative designs, evaluate retrieval performance against annotated benchmarks, incorporate account-level metadata, compare crisis and non-crisis baselines, and develop robust spam-filtering standards for federated platforms.

Future research should extend this framework in six directions. First, temporal lag analysis could determine whether X-Twitter bottleneck formation precedes or follows peaks in Bluesky commentary, clarifying causal ordering in cross-platform cascade dynamics. Second, semantic drift could be measured with embedding models such as SBERT or word2vec to capture how executive seed posts mutate as they move across platforms. Third, systematic network analysis using ERGM or centrality measures could identify elite proxy influence more rigorously and test whether proxy dominance is predictable from pre-crisis network position. Fourth, the metric-to-semantic linkage model should be expanded to additional environments, including Telegram, Truth Social's reply layer if it becomes accessible, and non-English platforms. Fifth, the governance implications of extreme cascade concentration warrant direct evaluation, especially whether interventions such as labeling coordinated amplification, slowing retweet velocity, or elevating original-text responses reduce bottleneck formation during crises. Finally, within-actor non-crisis baseline comparisons would help determine whether the divergences documented here are specific to military crisis communication or reflect more stable properties of platform amplification infrastructures.

**Acknowledgments:** The author(s) received no assistance that warrants acknowledgment.

**Declaration of Conflicting Interests:** The author(s) declared no potential conflicts of interest with respect to the research, authorship, and/or publication of this article.

**Funding:** The author(s) received no financial support for the research, authorship, and/or publication of this article.

**Ethical Approval and Informed Consent:** Ethical approval was not required for this study because it relied exclusively on publicly available/secondary, aggregated data and did not involve human participants.

**Data Availability Statement:** The data supporting the findings of this study are publicly available in the figshare repository: Mehmetcik, Hakan (2026). *Replication data for Elite Proxies, Algorithmic Bottlenecks, and Multi-Platform Information Cascades in the 2026 Iran War*. figshare. Dataset. https://doi.org/10.6084/m9.figshare.32589120.v2.